# The *Fermi Gamma Ray Space Telescope* discovers the Pulsar in the Young Galactic Supernova-Remnant CTA 1

A. A. Abdo[1,2], M. Ackermann[3], W. B. Atwood[4], L. Baldini[5], J. Ballet[6], G. Barbiellini[9,10], M. G. Baring[11], D. Bastieri[12,13], B. M. Baughman[14], K. Bechtol[3], R. Bellazzini[5], B. Berenji[3], R. D. Blandford[3], E. D. Bloom[3], G. Bogaert[15], E. Bonamente[16,17], A. W. Borgland[3], J. Bregeon[5], A. Brez[5], M. Brigida[18,19], P. Bruel[15], T. H. Burnett[20], G. A. Caliandro[18,19], R. A. Cameron[3], P. A. Caraveo[21], P. Carlson[22], J. M. Casandjian[6], C. Cecchi[16,17], E. Charles[3], A. Chekhtman[23,2], C. C. Cheung[8], J. Chiang[3], S. Ciprini[16,17], R. Claus[3], J. Cohen-Tanugi[24], L. R. Cominsky[25], J. Conrad[22,26], S. Cutini[27], D. S. Davis[8,28], C. D. Dermer[2], A. de Angelis[29], F. de Palma[18,19], S. W. Digel[3], M. Dormody[4], E. do Couto e Silva[3], P. S. Drell[3], R. Dubois[3], D. Dumora[30,31], Y. Edmonds[3], C. Farnier[24], W. B. Focke[3], Y. Fukazawa[32], S. Funk[3], P. Fusco[18,19], F. Gargano[19], D. Gasparrini[27], N. Gehrels[8,33], S. Germani[16,17], B. Giebels[15], N. Giglietto[18,19], F. Giordano[18,19], T. Glanzman[3], G. Godfrey[3], I. A. Grenier[6], M.-H. Grondin[30,31], J. E. Grove[2], L. Guillemot[30,31], S. Guiriec[24], A. K. Harding[8], R. C. Hartman[8], E. Hays[8], R. E. Hughes[14], G. Jóhannesson[3], A. S. Johnson[3], R. P. Johnson[4], T. J. Johnson[8,33], W. N. Johnson[2], T. Kamae[3], Y. Kanai[34], G. Kanbach[35], H. Katagiri[32], N. Kawai[36,34], M. Kerr[20], T. Kishishita[37], B. Kızıltan[38], J. Knödlseder[39], M. L. Kocian[3], N. Komin[6*,24], F. Kuehn[14], M. Kuss[5], L. Latronico[5], M. Lemoine-Goumard[30,31], F. Longo[9,10], V. Lonjou[30,31], F. Loparco[18,19], B. Lott[30,31], M. N. Lovellette[2], P. Lubrano[16,17], M. Marelli[21], M. N. Mazziotta[19], J. E. McEnery[8], S. McGlynn[22], C. Meurer[26], P. F. Michelson[3], T. Mineo[40], W. Mitthumsiri[3], T. Mizuno[32], A. A. Moiseev[7], C. Monte[18,19], M. E. Monzani[3], A. Morselli[41], I. V. Moskalenko[3], S. Murgia[3], T. Nakamori[34], P. L. Nolan[3], E. Nuss[24], M. Ohno[37], T. Ohsugi[32], A. Okumura[42], N. Omodei[5], E. Orlando[35], J. F. Ormes[43], M. Ozaki[37], D. Paneque[3], J. H. Panetta[3], D. Parent[30,31], V. Pelassa[24], M. Pepe[16,17], M. Pesce-Rollins[5], G. Piano[41], L. Pieri[12], F. Piron[24], T. A. Porter[4], S. Rainò[18,19], R. Rando[12,13], P. S. Ray[2], M. Razzano[5], A. Reimer[3], O. Reimer[3], T. Reposeur[30,31], S. Ritz[8,33], L. S. Rochester[3], A. Y. Rodriguez[44], R. W. Romani[3], M. Roth[20], F. Ryde[22], H. F.-W. Sadrozinski[4], D. Sanchez[15], A. Sander[14], P. M. Saz Parkinson[4], T. L. Schalk[4], A. Sellerholm[26], C. Sgrò[5], E. J. Siskind[45], D. A. Smith[30,31], P. D. Smith[14], G. Spandre[5], P. Spinelli[18,19], J.-L. Starck[6], M. S. Strickman[2], D. J. Suson[46], H. Tajima[3], H. Takahashi[32], T. Takahashi[37], T. Tanaka[3], J. B. Thayer[3], J. G. Thayer[3], D. J. Thompson[8], S. E. Thorsett[4], L. Tibaldo[12,13], D. F. Torres[47,44], G. Tosti[16,17], A. Tramacere[48,3], T. L. Usher[3], A. Van Etten[3], N. Vilchez[39], V. Vitale[41], P. Wang[3], K. Watters[3], B. L. Winer[14], K. S. Wood[2], H. Yasuda[32], T. Ylinen[49,22], M. Ziegler[4]

Contact authors: G. Kanbach[35], K. Wood[2], M. Ziegler[4]

[1]National Research Council Research Associate
[2]Space Science Division, Naval Research Laboratory, Washington, DC 20375
[3]W. W. Hansen Experimental Physics Laboratory, Kavli Institute for Particle Astrophysics and Cosmology, Department of Physics and Stanford Linear Accelerator Center, Stanford University, Stanford, CA 94305
[4]Santa Cruz Institute for Particle Physics, Department of Physics and Department of Astronomy and Astrophysics, University of California at Santa Cruz, Santa Cruz, CA 95064
[5]Istituto Nazionale di Fisica Nucleare, Sezione di Pisa, I-56127 Pisa, Italy
[6]Laboratoire AIM, CEA-CNRS-Université Paris Diderot, Service d'Astrophysique, CEA Saclay, F-91191 Gif sur Yvette, France
[7]Center for Research and Exploration in Space Science and Technology (CRESST), NASA Goddard Space Flight Center, Greenbelt, MD 20771
[8]NASA Goddard Space Flight Center, Greenbelt, MD 20771
[9]Istituto Nazionale di Fisica Nucleare, Sezione di Trieste, I-34127 Trieste, Italy
[10]Dipartimento di Fisica, Università di Trieste, I-34127 Trieste, Italy
[11]Rice University, Department of Physics and Astronomy, MS-108, P. O. Box 1892, Houston, TX 77251, USA
[12]Istituto Nazionale di Fisica Nucleare, Sezione di Padova, I-35131 Padova, Italy
[13]Dipartimento di Fisica “G. Galilei”, Università di Padova, I-35131 Padova, Italy

[14]Department of Physics, Center for Cosmology and Astro-Particle Physics, The Ohio State University, Columbus, OH 43210
[15]Laboratoire Leprince-Ringuet, École polytechnique, CNRS/IN2P3, Palaiseau, France
[16]Istituto Nazionale di Fisica Nucleare, Sezione di Perugia, I-06123 Perugia, Italy
[17]Dipartimento di Fisica, Università degli Studi di Perugia, I-06123 Perugia, Italy
[18]Dipartimento di Fisica "M. Merlin" dell'Università e del Politecnico di Bari, I-70126 Bari, Italy
[19]Istituto Nazionale di Fisica Nucleare, Sezione di Bari, 70126 Bari, Italy
[20]Department of Physics, University of Washington, Seattle, WA 98195-1560
[21]INAF-Istituto di Astrofisica Spaziale e Fisica Cosmica, I-20133 Milano, Italy
[22]Department of Physics, Royal Institute of Technology (KTH), AlbaNova, SE-106 91 Stockholm, Sweden
[23]George Mason University, Fairfax, VA 22030
[24]Laboratoire de Physique Théorique et Astroparticules, Université Montpellier 2, CNRS/IN2P3, Montpellier, France
[25]Department of Physics and Astronomy, Sonoma State University, Rohnert Park, CA 94928-3609
[26]Department of Physics, Stockholm University, AlbaNova, SE-106 91 Stockholm, Sweden
[27]ASI Science Data Center, I-00044 Frascati (Roma), Italy
[28]Center for Space Sciences and Technology, University of Maryland, Baltimore County, Baltimore, MD 21250, USA
[29]Dipartimento di Fisica, Università di Udine and Istituto Nazionale di Fisica Nucleare, Sezione di Trieste, Gruppo Collegato di Udine, I-33100 Udine, Italy
[30]CNRS/IN2P3, Centre d'Études Nucléaires Bordeaux Gradignan, UMR 5797, Gradignan, 33175, France
[31]Université de Bordeaux, Centre d'Études Nucléaires Bordeaux Gradignan, UMR 5797, Gradignan, 33175, France
[32]Department of Physical Science and Hiroshima Astrophysical Science Center, Hiroshima University, Higashi-Hiroshima 739-8526, Japan
[33]University of Maryland, College Park, MD 20742
[34]Department of Physics, Tokyo Institute of Technology, Meguro-ku, Tokyo 152-8551, Japan
[35]Max-Planck-Institut für Extraterrestrische Physik, Giessenbachstraße, 85748 Garching, Germany
[36]Cosmic Radiation Laboratory, Institute of Physical and Chemical Research (RIKEN), Wako, Saitama 351-0198, Japan
[37]Institute of Space and Astronautical Science, JAXA, 3-1-1 Yoshinodai, Sagamihara, Kanagawa 229-8510, Japan
[38]UCO/Lick Observatories, 1156 High Street, Santa Cruz, CA 95064, USA
[39]Centre d'Étude Spatiale des Rayonnements, CNRS/UPS, BP 44346, F-30128 Toulouse Cedex 4, France
[40]IASF Palermo, 90146 Palermo, Italy
[41]Istituto Nazionale di Fisica Nucleare, Sezione di Roma "Tor Vergata" and Dipartimento di Fisica, Università di Roma "Tor Vergata", I-00133 Roma, Italy
[42]Department of Physics, Graduate School of Science, University of Tokyo, 7-3-1 Hongo, Bunkyo-ku, Tokyo 113-0033, Japan
[43]Department of Physics and Astronomy, University of Denver, Denver, CO 80208
[44]Institut de Ciencies de l'Espai (IEEC-CSIC), Campus UAB, 08193 Barcelona, Spain
[45]NYCB Real-Time Computing Inc., 18 Meudon Drive, Lattingtown, NY 11560-1025
[46]Department of Chemistry and Physics, Purdue University Calumet, Hammond, IN 46323-2094
[47]Institució Catalana de Recerca i Estudis Avançats (ICREA), Barcelona, Spain
[48]Consorzio Interuniversitario per la Fisica Spaziale (CIFS), I-10133 Torino, Italy
[49]School of Pure and Applied Natural Sciences, University of Kalmar, SE-391 82 Kalmar, Sweden
* current address

**Abstract:**

Energetic young pulsars and expanding blast waves (supernova remnants, SNRs) are the most visible remains after massive stars, ending their lives, explode in core-collapse supernovae. The *Fermi Gamma-Ray Space Telescope* has unveiled a radio quiet pulsar located near the center of the compact synchrotron nebula inside the supernova remnant CTA 1. The pulsar, discovered through its gamma-ray pulsations, has a period of 316.86 ms, a period derivative of 3.614 x $10^{-13}$ s $s^{-1}$ . Its characteristic age of $10^4$ years is comparable to that estimated for the SNR. It is conjectured that most unidentified Galactic gamma ray sources associated with star-forming regions and SNRs are such young pulsars.

---

After the discovery of radio pulsars in the late 1960's some clear SNR-pulsar associations were discovered, e.g. most notably the Crab and Vela systems. Observations in the radio, X-ray and gamma-ray bands with increasing sensitivity during the last 30 years have added many more SNR-pulsar associations. But we are still far from the complete census of these products of massive star deaths, which is needed to study a major population of stellar and Galactic astronomy. Here we report the discovery of a gamma-ray pulsar with spin period 316 ms, coinciding with the previously known gamma-ray source 3EG J0010+7309, thus confirming the identification of the neutron star powering the PWN and the gamma-ray source. This pulsar detection implies that many of the yet-unidentified low latitude Galactic gamma-ray sources also could be pulsars.

A survey of radio sources conducted at 960 MHz with the Owens Valley Radio Observatory in the 1960's discovered a previously uncataloged extended source (*1*), designated CTA 1, as the first object in Caltech's catalogue A. Follow-up radio surveys (*2-7*) with increased sensitivity and angular resolution showed that CTA 1 has the typical morphology of a shell-type SNR with an incomplete shell of filaments and extended emission from a broken shell roughly circular and about 90' diameter (Fig. 1). The excitation of atomic lines in the shocked ISM with well-defined optical filaments (*8*) lends further support to the identification of CTA 1 as a young SNR in the Sedov phase of expansion. The radio and X-ray characteristics of CTA 1 imply that it is 1.4±0.3 kpc away (*6*) and it exploded 5000 to 15000 years ago (*6,10,11*).

Imaging and spectroscopy of CTA 1 with *ROSAT* (*9*) , *ASCA* (*10*) , *XMM* (*11*) , and *CHANDRA* (*12*) revealed a typical center-filled or composite SNR with a central point source, RXJ0007.0+7303, embedded in a compact nebula, and a jet-like extension (*12*). The offset of the X-ray source from the geometrical center of the SNR suggested that it has a transverse velocity (*11*) of ~450 km $s^{-1}$. The natural interpretation of these data is that of a young neutron star (NS), visible both in thermal surface and non-thermal magnetospheric emission (*12*), powering a synchrotron pulsar wind nebula (PWN). The thermal spectrum from the NS (*11,12*) is not easily interpreted: the temperature is too high and the required emission area is too small if the NS has no atmosphere. A particle-heated polar cap could be a possibility. Alternatively, if the NS has a light element atmosphere and cools through a direct URCA process, a cooling age of (1-2)×$10^4$ years is also possible (*11*). Although no signs of periodicity could be found in the X-ray data (*11*) the energetics of the PWN lead to typical requirements for the time-averaged spin-down power of the putative pulsar of $10^{36}$ – $10^{37}$ erg/s. Very deep searches (*12*) for a counterpart of the X-ray source in radio and optical wavebands resulted only in upper limits. If RX J0007.0+7303 is indeed a radio pulsar, its radio luminosity is an order of magnitude below the faintest radio pulsars known (*12*). It is likely that the radio beam does not intersect the Earth.

High-energy emission (>100 MeV) from the EGRET source 3EG J0010+7309 matches RX J0007.0+7303 spatially although the EGRET position uncertainty is very large (13). The position derived from EGRET photons above 1 GeV (14) at l,b = 119.87, 10.52 with an error radius of 11 arcmin (95% confidence) overlaps even better with the ROSAT source. It has thus been suggested that the 3EG source is an unresolved source in CTA 1 (e.g. 14), or, more generally, that several unidentified gamma-ray sources are associated with SNRs (e.g. 15). Confirmation of such SNR associations based on imaging was however not possible with the *EGRET* angular resolution. The CTA 1 gamma-ray source shows all indications of being a young pulsar: the gamma-ray flux was constant through the epochs of *EGRET* observations (1991-95) and the spectrum showed a hard power law with an index of -1.6±0.2 and a spectral steepening above ~2 GeV (*14*), which is similar to other *EGRET* pulsars like Geminga and Vela. For an assumed pulsar beam of 1 sr, and taking into account the uncertainty in distance, the observed gamma-ray flux corresponds to a luminosity of $(4\pm2)\times10^{33}$ erg/s, well within the range of the luminosities of Geminga ($9\times10^{32}$ erg/s, P~237ms ) and Crab ($4\times10^{34}$ erg/s, P~33ms).

On 11 June 2008, the *Fermi Gamma-ray Space Telescope* was launched into a low earth orbit (*16*). The imaging gamma-ray telescope LAT (Large Area Telescope), *Fermi*'s main instrument, covers the energy range from 20 MeV up to greater than 300 GeV with a sensitivity that exceeds that of *EGRET*. The first exposures of the region of CTA 1 were made during the commissioning phase of *Fermi* LAT (30 June to 30 July 2008) and in the initial days (5-20 August 2008) of routine operations. Although the telescope was not yet fully tuned and calibrated during commissioning, more than 900 gamma-ray photons above 100 MeV from 3EG J0010+7309 were recorded during these exposures (31), which amounts to about 2.6 times the number collected with *EGRET* from this source over its entire mission.

A bright gamma-ray source is detected at $l,b$ = 119°.652, 10°.468 with a 95% (statistical) error circle radius of 0°.038 (a systematic error of about 0°.02 is not included). Fig. 1 shows the LAT source and the X-ray source RX J00070+7302, which is located central to the PWN, superimposed on the radio map at 1420 MHz (7). These fall on the edge of 3EG J0010+7309 ($l,b$ = 119.92, 10.54) and its 95% error circle of radius 0.24°. The measured flux of the LAT source is $(3.8\pm0.2)\times10^{-7}$ ph (>100 MeV) $cm^{-2}$ $s^{-1}$, with an additional systematic uncertainty of 30% owing to the ongoing calibration of the instrument, which is consistent with the *EGRET* measured flux of $(4.2\pm0.5) \times 10^{-7}$ ph $cm^{-2}$ $s^{-1}$ (*13*).

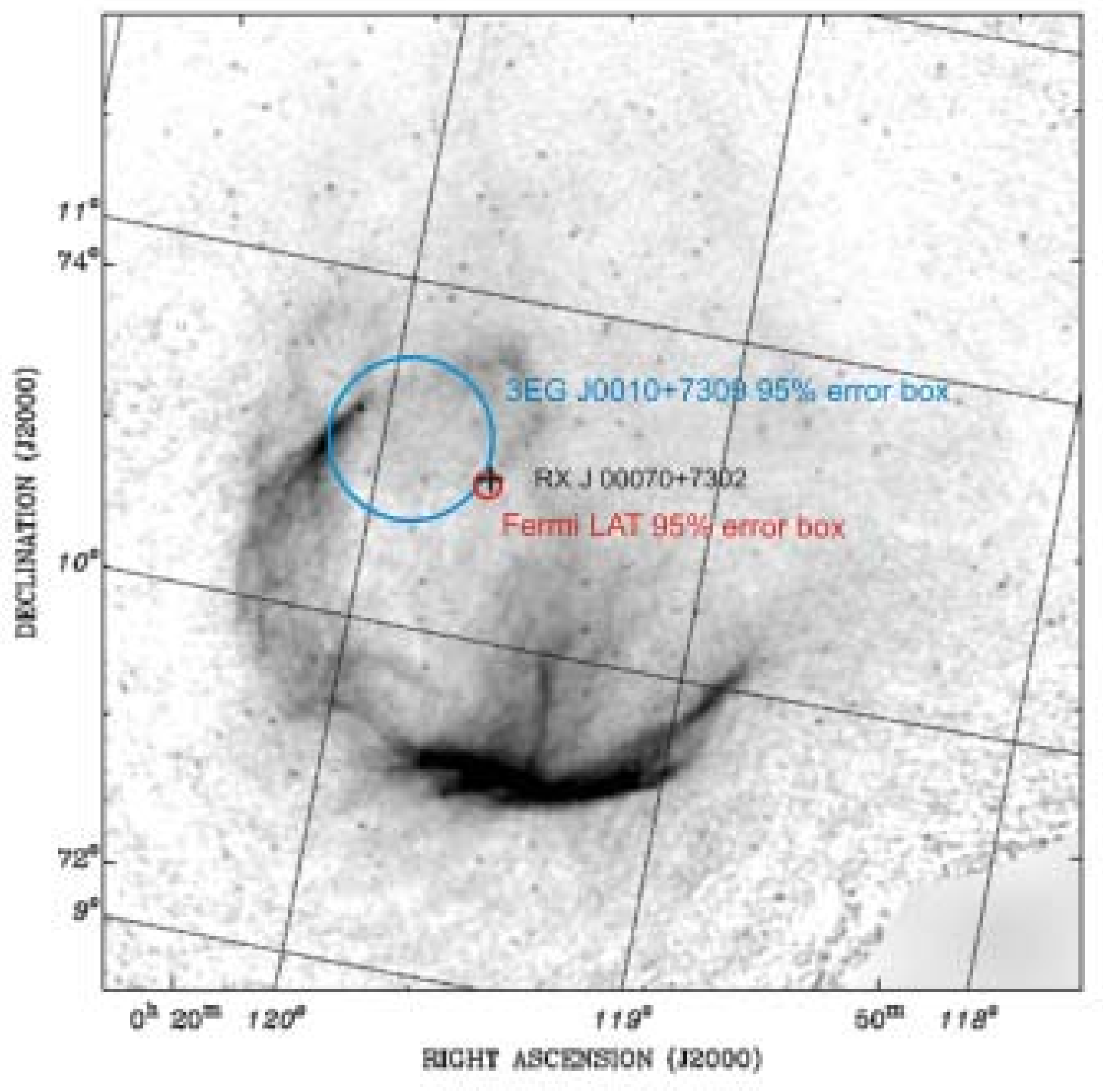

**Fig. 1:** The *Fermi* LAT gamma-ray source, the central PWN X-ray source, and the corresponding *EGRET* source superimposed on a 1420 MHz map (*7*) of CTA 1.
The LAT source and its 95% error region (small red circle) is displayed on the map together with the central PWN source RX J00070+7302 (cross) and the position and error of the corresponding *EGRET* source 3EG J0010+7309 (large blue circle). The coincidence of the pulsed gamma-ray source and the X-ray point source embedded in the off-center PWN is striking. The offset of the pulsar from the center of the radio SNR, which is thought to be the place of origin, is quite visible. The inferred transverse speed of the pulsar is ~450 km/s, which is a reasonable speed of a pulsar (*20*).

The arrival times of the LAT photons, which are recorded with 300 ns accuracy and referenced to the *Fermi* satellite GPS clock, were corrected to the solar-system barycenter (SSB) using the JPL DE405 solar system ephemeris and the *Chandra* X-ray position (see table 1). Photons with energy greater than 100 MeV were selected within a radius of 1° around the source position and searched for periodicity. The results described here are not changed significantly if photons within selection radii between 0.7° and 2.5° are used. Application of a new search technique based on photon arrival time differencing (*17*), which is highly efficient for sparse photon data, and refining and fitting the detections with the pulsar analysis packages PRESTO (*18*) and Tempo2 (*19*) resulted in the detection of significant pulsations in the selected photons (31). Fig. S1 in the supporting online material shows that the pulsations are significantly present over the complete time interval of observation. The pulsar rotational ephemeris is given in table 1.

| | |
|---|---|
| Frequency (Hz) | 3.165922467(9) |
| Frequency derivative ($s^{-2}$) | -3.623(4) $\times 10^{-12}$ |
| Period (ms) | 315.8637050(9) |
| Period derivative ($s\ s^{-1}$) | 3.615(4) $\times 10^{-13}$ |
| Epoch (MJD (TDB)) | 54647.440 938 |
| R.A. (J2000.0) | $00^h\ 07^m\ 01^s.56$ |
| DEC. (J2000.0) | +73° 03´ 08´´.1 |
| Galactic longitude | 119°.65947(3) |
| Galactic latitude | +10°.463348(3) |

**Tab. 1:** Rotational ephemeris for the pulsar in CTA 1. The numbers in parentheses indicate the error in the last decimal digit.  For the SSB correction the position of the *Chandra* X-ray source (*12*) was assumed.

Extracting photons around the Vela pulsar from the same set of observations and applying the same analysis procedures, the rotational ephemeris for the Vela pulsar was found to be in good agreement with the values obtained by the LAT radio pulsar timing collaboration (30).

A contour plot of period-period derivative search space reveals the pulsar (Fig. 2) as does the resulting gamma-ray light- curve above 100 MeV (Fig. 3).

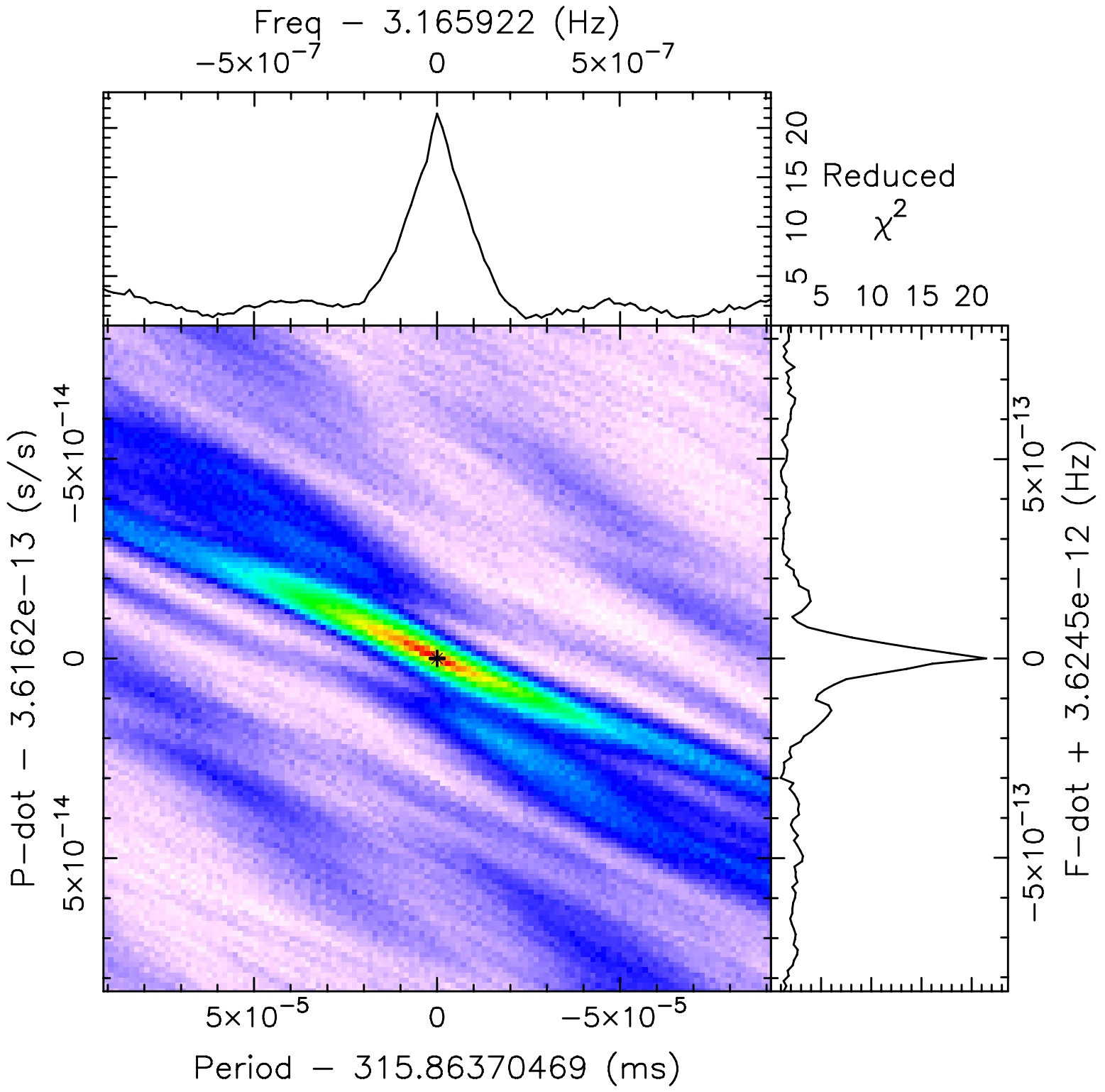

**Fig. 2:** Contours of detection significance over a range of period and period derivative using photons within a radius of 1° around RXJ0007.0+7303. The initial indication of a signal in this $P$, $\dot{P}$ region was found with a novel search technique using photon arrival time differencing (*17*), while the determination of the exact ephemeris makes use of the tools PRESTO and Tempo2 (*18, 19*)

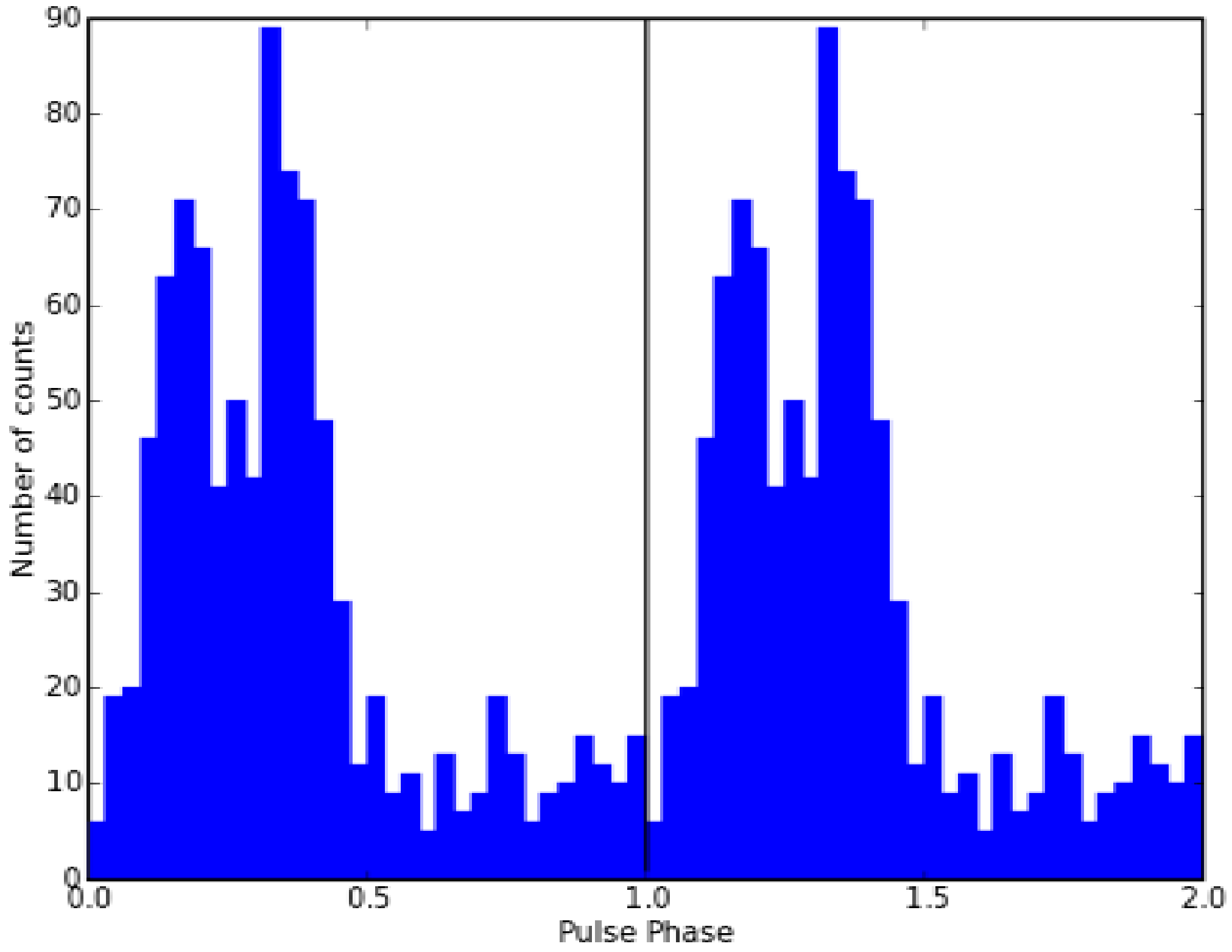

**Fig. 3:** Gamma-ray (>100 MeV) light-curve of the pulsar in CTA 1 shown over two periods of rotation with a resolution of 32 phase bins per period (corresponding to ~10 ms/bin). The two maxima in the broad emission feature each have a FWHM of ~0.12 and are separated by about 0.2 in phase. Overall, the LAT pulsar light-curve is similar to the gamma-ray light-curve of the *EGRET* pulsar PSR B1706-44 (*21*).

A neutron star with a moment of inertia of $I$ =1.0 $\times 10^{45}$ g cm$^2$ and angular frequency $\omega$ is assumed to lose its rotational energy through magnetic dipole radiation and follow a braking law of $\dot{\omega} \propto -\omega^3$. This can be integrated to yield the characteristic age $\tau = \omega / 2\dot{\omega}$, which is a coarse estimate of the true age of a pulsar. The spin-down power $\dot{E}_{rot} = I\omega\dot{\omega}$ and the dipole magnetic field strength, $B = 3.2 \times 10^{19} \sqrt{P\dot{P}}$ G, also follow from the parameters of rotation.

For the CTA 1 pulsar we derive a characteristic age of ~$1.4\times10^4$ years, a spin-down power of ~$4.5\times10^{35}$ erg s$^{-1}$, and a surface magnetic field strength of $1.1\times10^{13}$ G. This field strength is higher than any of the *EGRET* detected pulsars and second highest among known gamma-ray pulsars. PSR J1509-58 with an inferred field of $1.54\times10^{13}$ G shows emission only up to ~30MeV, while emission from the CTA 1 pulsar is present to at least 5 GeV.

We searched archival data of exposures by *XMM*, *ASCA*, *Chandra*, and *EGRET* for periods near that extrapolated from the LAT ephemeris. The pulsar was not significantly detected in these data (*22*).

The new pulsar in CTA 1 exhibits all the characteristics of a young high-energy pulsar, which powers a synchrotron pulsar wind nebula embedded in a larger SNR. The spin-down power of the CTA 1 pulsar of ~$4.5\times10^{35}$ erg s$^{-1}$ is sufficient to supply the PWN with magnetic fields and energetic electrons at the required rate of $10^{35}$-$10^{36}$ erg s$^{-1}$ (*11*) and the pulsar age is consistent with the inferred age for the SNR. The Crab pulsar with its spin-down power of ~$4.5\times10^{38}$ erg $s^{-1}$ supplies the Crab nebula with its requirement of ~$10^{38}$ erg $s^{-1}$ with a similar efficiency (*23*). Comparing the total luminosity from the CTA 1 pulsar inferred from the LAT flux measured for 3EG J0010+7309 to the pulsar spin-down power, we estimate an efficiency of converting spin-down power into pulsed gamma rays that is about 1%, if the emission is beamed into a solid angle of 1 sr. The Vela pulsar, which is of similar age, has a gamma-ray efficiency of 0.1%; however PSR B1706-44 (*21*) with an age of $1.7\times10^4$ years has an efficiency of 1.9%, and the much older Geminga pulsar ($3.4\times10^5$ years) converts its spin-down into energetic gamma-rays with an efficiency of about 3%.

The gamma-ray characteristics and the absence of a radio signal (*12*) from the CTA 1 pulsar is suggestive of the family of outer-gap or slot-gap model descriptions for energetic pulsars (*24, 25, 26*). These outer-magnetosphere models naturally generate gamma-ray emission over a broad range of phase, with superposed sharp peaks resulting from caustics in the pattern of the emitted radiation. Since the emission is predicted to cover such a large area of the sky (>> 1 sr) in such models, the total radiated luminosity as inferred from the observed pulsed flux could result in an efficiency as high as 10%, depending on the magnetic inclination angle. The absence of the radio signal is readily explained by misalignment of a narrow radio beam and our line of sight. Both conditions can be met if we see the pulsar at a large angle with respect to the spin and magnetic field axes, but only a detailed model can quantify the viewing geometry of the CTA 1 pulsar. Spectral cut-offs at energies of 1-10 GeV, that would be indicative of a polar cap mechanism with attenuation of outgoing photons via magnetic pair creation (*27*) or curvature radiation-reaction limited acceleration in an outer gap (*28*) are not yet discernible.

Our detection of a new gamma-ray pulsar during the initial operation of the *Fermi* LAT implies that there may be many gamma-ray-loud but X-ray- and radio-quiet pulsars. Although 3EG J0010+7309 was long suspected to be a gamma-ray pulsar because of its clear association with a SNR at about 10° Galactic latitude, it is also fairly typical of many unidentified *EGRET* sources. If the radio remnant CTA 1 were located at low Galactic latitudes it could have been more difficult to recognize because of the higher and structured radio background of the Galactic disk. About 75% of the low Galactic latitude *EGRET* sources ($|b|<10°$, about 100 objects in the 3EG catalog and CTA 1 just on the edge of this region) are still unidentified, although several are associated with SNRs.

The unidentified low Galactic latitude *EGRET* sources represent the closer and brighter objects of a Galaxy wide population of gamma-ray sources. *EGRET* was not sensitive enough to discern the more distant sources, which blurred into the diffuse Galactic emission. A model Galactic population conforming to the *EGRET* measurements (*29*), distributed in galacto-centric distance and height above the disk, yields several thousand sources, of typical >100MeV luminosities in the range $6\times10^{34}$ - $4\times10^{35}$ erg/s. The CTA 1 pulsar with an isotropic luminosity above 100 MeV of ~$6\times10^{34}$ erg/s falls in the range required for these sources. The CTA 1 pulsar detection implies that gamma-loud but radio and X-ray faint pulsars are likely to be detectable in a fair fraction of the remnants of massive star deaths. Topics as diverse as the SN rate in the Galaxy, the development of young (including historical) SNRs and the physics of pulsar emission can then be studied.

**Acknowledgments:**

The Fermi LAT Collaboration acknowledges the generous support of a number of agencies and institutes that have supported the Fermi LAT Collaboration. These include the National Aeronautics and Space Administration and the Department of Energy in the United States, the Commissariat à l'Energie Atomique and the Centre National de la Recherche Scientifique / Institut National de Physique Nucléaire et de Physique des Particules in France, the Agenzia Spaziale Italiana and the Istituto Nazionale di Fisica Nucleare in Italy, the Ministry of Education, Culture, Sports, Science and Technology (MEXT), High Energy Accelerator Research Organization (KEK) and Japan Aerospace Exploration Agency (JAXA) in Japan, and the K. A. Wallenberg Foundation and the Swedish National Space Board in Sweden.

**Supporting Online Material (to be stored in a public web-site)**

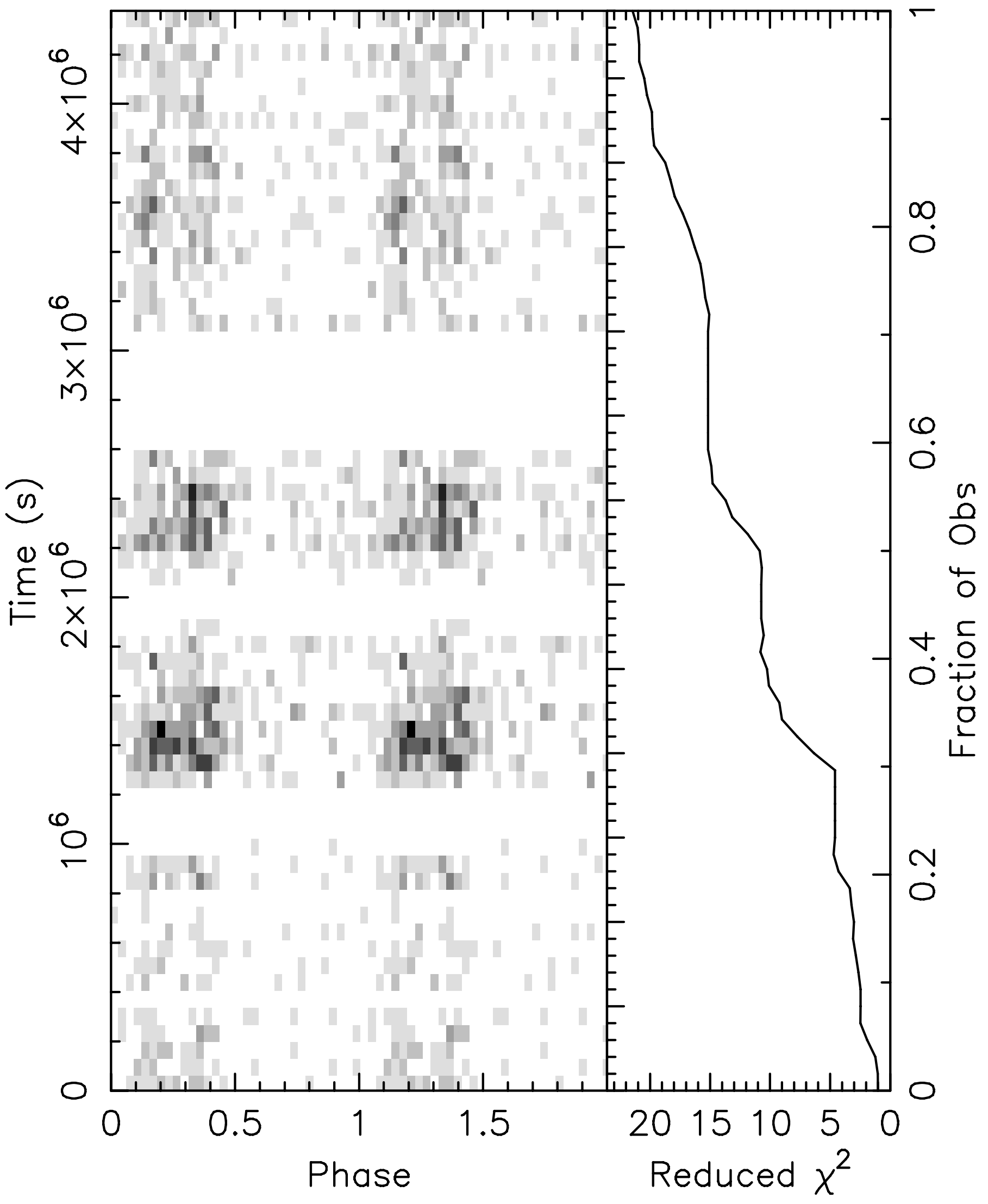

Fig. S1: Photon density (grey scale) as a function of pulsar phase and epoch of observations. The origin of the time axis is 2008-06-30 09:40:35. The right subplot show how the significance of the detection (represented as the reduced chi^2) increases with time. The two time periods with the strongest detection (beginning at 1.2E6 and 2.2E6

seconds into the observation) are the result of increased time on source during 3-axis pointed observations that were part of the Fermi Launch & Early Operations testing. The rest of the data were taken in the nominal sky survey mode.